\documentclass[aps,pre,floats,twocolumn,showpacs,superscriptaddress]{revtex4}

\usepackage{graphicx,epsfig}
\usepackage{times}
\usepackage{amssymb,amsmath,multirow,rotate,color}
\usepackage{xcolor}

\begin{document}

\title{Markovian approach to tackle the interaction of simultaneous diseases}

\author{D. Soriano-Pa\~nos} 
\affiliation{GOTHAM lab., Institute for Biocomputation and Physics of Complex Systems (BIFI), University of Zaragoza, 50018 Zaragoza, Spain}  
\affiliation{Departamento de F{\'i}sica de la Materia Condensada, Universidad de Zaragoza, 50009 Zaragoza, Spain}
\author{F. Ghanbarnejad}
\affiliation{Institute of Theoretical Physics, Technical University of Berlin, Hardenbergstr. 36, Sekr. EW 7-1 D-10623 Berlin}
\affiliation{Quantitative Life Sciences (QLS), The Abdus Salam International Centre for Theoretical Physics (ICTP), 34151 Trieste , Italy}
\author{S. Meloni}
\affiliation{IFISC,  Instituto de F\'{\i}sica Interdisciplinar y Sistemas Complejos (CSIC-UIB), 07122 Palma de Mallorca, Spain}
\author{J. G{\'o}mez-Garde\~{n}es}\email{gardenes@unizar.es}
\affiliation{GOTHAM lab., Institute for Biocomputation and Physics of Complex Systems (BIFI), University of Zaragoza, 50018 Zaragoza, Spain}  
\affiliation{Departamento de F{\'i}sica de la Materia Condensada, Universidad de Zaragoza, 50009 Zaragoza, Spain}
\date{\today}

\begin{abstract}
The simultaneous emergence of several abrupt disease outbreaks or the extinction of some serotypes of multi-strain diseases are fingerprints of the interaction between pathogens spreading within the same population. Here, we propose a general and versatile benchmark to address the unfolding of both cooperative and competitive interacting diseases. We characterize the explosive transitions between the disease-free and the epidemic regimes arising from the cooperation between pathogens, and show the critical degree of cooperation needed for the onset of such abrupt transitions. For the competing diseases, we characterize the mutually exclusive case and derive analytically the transition point between the full-dominance phase, in which only one pathogen propagates, and the coexistence regime. Finally, we use this framework to analyze the behavior of the former transition point as the competition between pathogens is relaxed.  \end{abstract}


\maketitle
\section{INTRODUCTION}

Containing the spread of infectious diseases constitutes one of the major challenges of modern societies \cite{WHO2016}. During the last decades, empowered with tools from non-equilibrium statistical physics and nonlinear dynamics, classical epidemic models \cite{epirep0,epirep1,epirep2,epirep3} have been progressively refined to capture the many ingredients that interplay to give rise to real epidemic outbreaks \cite{epirep4,epirep5,epirep6,epirep7}. 
Epidemic modeling was particularly boosted by the incorporation of complex interaction networks \cite{net2,net3,net4}: capturing the backbone of interactions through which pathogens spread allowed to study how network heterogeneity influences epidemic onsets \cite{Het1,Het2,Het3,Het4}. 
\smallskip

The advances made in the representation of the interaction map in a variety of complex systems (such as multilayer frameworks \cite{DeDomenicoPRX,Boccaletti,Kivela2014JCN} or time-varying graphs \cite{TVbook}) have allowed to  improve epidemic models to tackle the analysis about the role played by human interactions in contagion processes. Examples of the ingredients incorporated in epidemic models include: the volatile nature of human contacts \cite{TVep0,TVep1,TVep2,TVep3}, the multi-scale and recurrent mobility patterns in metapopulation models \cite{metapop0,metapop1,metapop2,metapop3,metapop4}, the coexistence of multiple interaction or mobility modes \cite{Mep1,Mep2,Mep3}, or the adaptation of human behavior to epidemic waves \cite{vacc1,vacc2,vacc3}. All these studies put the focus on understanding the role of the former ingredients to propose efficient containment policies \cite{quar1,quar2,quar3} aimed at controlling and preventing the unfolding of epidemic states.
\smallskip

The vast majority of these models are designed to characterize the spreading dynamics of single pathogens whose evolution is assumed not to depend on the presence of others. However, many diseases do not fulfil this assumption since their spreading patterns are strongly influenced by the simultaneous propagation of other pathogens. One paradigmatic example where the interaction between simultaneous diseases played a crucial role on their impact took place in 1918, manifested by a sudden abrupt increase in the death rate per Pneumonia cases matched with the onset of the Spanish Flu \cite{flupneu}. The correlation between both diseases clearly suggested their interplay in a cooperative way. On the other hand, there are other examples in which the presence of one pathogen is detrimental to the propagation of other infections, since being infected by one disease confers partial or total immunity with respect to the other one. Competitive interaction between diseases typically occurs within the different serotypes of multi-strain diseases such as DENV \cite{DENV} or Influenza \cite{Influenza}, but also occur between pathogens corresponding to different diseases, such as the recently reported interaction between ZIKV and DENV \cite{int11}.
\smallskip

In the recent years, several works \cite{int1,int2,int3,HDA,int4,int5,int11,int12,int6,int7,int8,int10} have tackled the extension of epidemic models in order to introduce the interaction between different co-existing diseases. Remarkably, interesting theoretical results have been already found such as the emergence of first-order transitions when two diseases cooperate \cite{int1,int2,int3,HDA,int4,int11,int12} or the extinction of some infectious strains due to the competition with their counterparts  \cite{int5,int6,int7,int8,int10}. However, some of these findings have been obtained at the expense of relying on strong assumptions regarding the contagion network structure or the mechanisms driving the interaction between pathogens; limiting their application to real scenarios where cooperative or competitive spread of pathogens occurs. In addition, many of the current models are specifically focused on understanding the effects of introducing either strong cooperation or mutually exclusive competition between two diseases. This way, the formulation of a consistent mathematical framework to characterize diseases whose interaction lies in between the former extreme cases remains as an open theoretical challenge.
\newpage

In this work, we propose a general epidemic model for characterizing the spread of interplaying pathogens with arbitrary degree of interaction, i.e., ranging from the mutually exclusive case to the strong cooperative regime. This new framework is constructed by following the Microscopic Markov Chain approach (MMCA) \cite{gomezEPL,gomezPRE}, so to keep all the information about the structure of the contact network without any statistical assumption. We show that the Markovian equations here proposed reproduce with great accuracy the variety of phenomena observed in Monte Carlo simulations, such as the explosive epidemic outbreaks in the cooperative case, thus providing an alternative to computationally expensive simulations. In addition, the Markovian framework opens the door to a reliable mathematical analysis of  the model. As an example, we derive analytically the second epidemic threshold that, in the fully-competitive case, separates the regions of full-dominance by one single pathogen and that of coexistence.  


\smallskip

The manuscript is organized as follows. In Section II we present the description of the model for interacting diseases and explain the mathematical formalism, including the theoretical assumptions and the rationale behind the equations governing the evolution of the system. In Section III and IV, we apply the theoretical framework to study cooperative and competitive dynamics respectively. In these sections we report and discuss the phenomena resulting from the interaction of the diseases, thus recovering, under a single framework, the most important findings observed in previous works. In both cases, we check the validity of the theoretical predictions by comparing them with results obtained from extensive Monte Carlo simulations. Finally, we round off the work in Section V by giving some conclusions derived from our model and discussing about its implication for future research.

\smallskip

\section{The model}
We start by assuming that contagion processes are dictated by an unweighted and undirected contact network of $N$ nodes, each accounting for an agent, whereas interactions are determined by the $L$ network links. The network is described by its adjacency matrix {\bf A} whose entries are defined as $A_{ij}=1$ if nodes $i$ and $j$ are connected and $A_{ij}=0$ otherwise. For the spreading dynamics, we consider that each disease $\alpha$ can be individually modeled as a Susceptible-Infected-Susceptible (SIS) dynamics, for which contagion and recovery probabilities are denoted as $p_{\alpha}$ and $r_{\alpha}$ respectively. In the absence of other pathogens, each disease $\alpha$ spreads from an infected agent to a susceptible one with probability $p_{\alpha}$, whereas infected agents become susceptible with probability $r_{\alpha}$. Here, for the sake of simplicity, we will restrict to the case of two interacting diseases so that $\alpha=1,2$. The interaction between the two diseases at work requires to couple two SIS dynamics. Therefore, agents subjected to a double SIS dynamics can be in four possible states which are: susceptible of contracting both diseases ($SS$), infected by the first disease and susceptible of the second ($IS$), susceptible of the first and infected of the second ($SI$), and infected by both pathogens ($II$). 

\begin{figure}[t!]
\centering
\includegraphics[width=0.95\columnwidth]{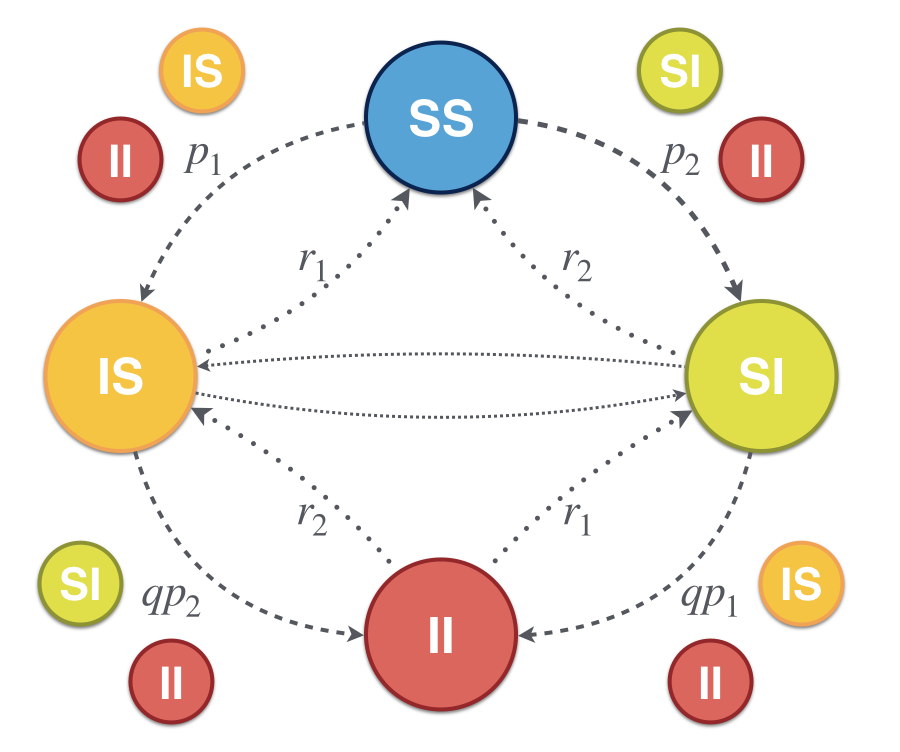}
\caption{Contagion and recovery microscopical processes considered in our model. Note that the contagion processes involving already infected individuals are influenced by the parameter $q$ which encodes the interaction between the two co-existing diseases.}
\label{fig1}
\end{figure}

\smallskip

We now describe the transitions that govern the two coupled SIS dynamics, i.e., we define the transition probabilities between the former four epidemic states. In Fig.~\ref{fig1} we show all the microscopical processes involved in our model. First, for healthy (SS) agents  we consider that the probability of being infected with pathogen $\alpha$ is not affected by the presence of the other one. Therefore, both pathogens are transmitted to $SS$ agents with probabilities $p_1$ and $p_2$ respectively. To apply our model to mutually exclusive diseases, double contagions of fully susceptible agents are forbidden. 
The interaction between both circulating diseases is included via a scaling parameter, $q$, affecting the probability that an agent already infected by one disease catches the other one, as shown in  Fig.~\ref{fig1} for the transitions $IS\rightarrow II$ and $SI\rightarrow II$. This way, $q<1$ implies that infected agents by one disease are less likely to get the other one, thus encoding a competition between both. On the other hand, $q>1$ means that being affected by one disease boosts the contagion by the other one, which corresponds to a cooperative regime.  Finally, all the recovery processes are assumed not to depend on the circulation of other pathogens, so that individuals overcome diseases $1$ and $2$ with probabilities $r_1$ and $r_2$ respectively. 

\smallskip

Mathematically, our formalism contains a set of interdependent Markovian equations which enable to track the temporal evolution of the dynamical state of each agent $i$. Since there are four possible epidemic states for each of the $N$ agents, we require $3N$ equations to completely characterize the evolution of the network. Given an agent, say $i$, let us denote as $[\rho^{\gamma}]_i^t$ the probability that this agent belongs to each of the following states $\gamma$ ($\gamma=IS,SI,II$) at time $t$. Under the microscopical rules defined above, the temporal evolution of these probabilities reads as follows:
\begin{widetext}
\begin{eqnarray}
[\rho^{II}]^{t+1}_{i} &=& [\rho^{SI}]_{i}^t (1-r_2)\left(1-\prod\limits_j^N\left[1-A_{ij}p_1q\left([\rho^{IS}]_{j}^t+[\rho^{II}]_{j}^t\right)\right]\right)  + [\rho^{IS}]_{i}^t (1-r_1)\left(1-\prod\limits_j^N\left[1-A_{ij}p_2q\left([\rho^{SI}]_{j}^t+[\rho^{II}]_{j}^t\right)\right]\right) \nonumber\\
&+&[\rho^{II}]_{i}^t\left(1-r_1\right)\left(1-r_2\right)\ ,\label{Eq.II}\\
& &  \nonumber \\
  \left[\rho^{IS}\right]^{t+1}_i &=& [\rho^{SI}]_i^t\left[r_2\left(1-\prod\limits_j^N\left[1-A_{ij}p_1q([\rho^{IS}]_j^t+[\rho^{II}]_j^t)\right]\right)\right] + [\rho^{IS}]_i^t\left(1-r_1\right)\prod\limits_j^N\left[1-A_{ij}p_2q([\rho^{SI}]_j^t+[\rho^{II}]_j^t)\right]\nonumber\\
 &+&[\rho^{II}]_i^t r_2(1-r_1) \nonumber \\ 
 &+&\left[\rho^{SS}\right]_i^t\left(1-\prod\limits_j^N\left[1-A_{ij}\left(p_1([\rho^{IS}]_j^t+[\rho^{II}]_j^t)+p_2([\rho^{SI}]_j^t+[\rho^{II}]_j^t)-p_1p_2([\rho^{II}]_j^t)^2\right)\right]\right)f_{IS}\ , \label{Eq.IS} \\
 \left[\rho^{SI}\right]^{t+1}_i &=& [\rho^{IS}]_i^t\left[r_1\left(1-\prod\limits_j^N\left[1-A_{ij}p_2q([\rho^{SI}]_j^t+[\rho^{II}]_j^t)\right]\right)\right] + [\rho^{SI}]_i^t\left(1-r_2\right)\prod\limits_j^N\left[1-A_{ij}p_1q([\rho^{IS}]_j^t+[\rho^{II}]_j^t)\right]\nonumber\\
 &+&[\rho^{II}]_i^t r_1(1-r_2) \nonumber \\ 
 &+&\left[\rho^{SS}\right]_i^t\left(1-\prod\limits_j^N\left[1-A_{ij}\left(p_1([\rho^{IS}]_j^t+[\rho^{II}]_j^t)+p_2([\rho^{SI}]_j^t+[\rho^{II}]_j^t)-p_1p_2([\rho^{II}]_j^t)^2\right)\right]\right)f_{SI}\ .\label{Eq.SI}
\end{eqnarray}
\end{widetext}
For the sake of readability, we have included the variable $\left[\rho^{SS}\right]_i^t$ whose value is automatically calculated as $\left[\rho^{SS}\right]_i^t= 1-\left[\rho^{IS}\right]_i^t-\left[\rho^{SI}\right]_i^t-\left[\rho^{II}\right]_i^t$. Note that the contagion processes involving totally susceptible (SS) agents are shaped by $f_{IS}$ and $f_{SI}$. These factors account for the probability of contracting one disease when exposed to the other pathogen as well. To define this probability, we must define a rule for the case in which a fully susceptible agent is in contact with both pathogens when interacting with its neighbors. In this scenario, we assume that each disease will be contracted with the same probability. This way, the probabilities $f_{IS}$ and $f_{SI}$ read as:
\begin{widetext}
\begin{eqnarray}
f_{IS} &=& \frac{g_{IS}\left(1-0.5g_{SI}\right)}{g_{IS}\left(1-0.5 g_{SI}\right)+g_{SI}\left(1-0.5 g_{IS}\right)}\;,\\
f_{SI} &=& \frac{g_{SI}\left(1-0.5g_{IS}\right)}{g_{IS}\left(1-0.5g_{SI}\right)+g_{SI}\left(1-0.5 g_{IS}\right)}\;,
\end{eqnarray}
\end{widetext}
where $g_{IS}$ and $g_{SI}$ are the probabilities of making at least one infectious contact with individuals affected by the first and the second disease respectively. These two probabilities can be expressed as:
\begin{eqnarray}
g_{IS}&=&1-\prod\limits_j^N\left[1-A_{ij}p_1\left([\rho^{IS}]_j^t+[\rho^{II}]_j^t\right)\right] \\
g_{SI}&=&1-\prod\limits_j^N\left[1-A_{ij}p_2\left([\rho^{SI}]_j^t+[\rho^{II}]_j^t\right)\right] 
\end{eqnarray}
With these two equations we complete the Markovian description for two interacting diseases provided by Eqs.~(1)-(3).
\bigskip

In the following sections we apply the formalism proposed above to study the impact of the interaction between simultaneous diseases. In particular, to put our focus on the effects of the interaction between diseases, we assume that both circulating pathogens, though different, are epidemiologically equivalent. Therefore, the only epidemiological parameters to be included in our framework are the infectivity $p=p_1=p_2$, the recovery rate, $r=r_1=r_2$, and the degree of interaction between pathogens, $q$. Regarding the contagion network, in the following sections we consider that our system is composed by $N=1000$ interacting agents whose contacts are governed by Erd\"os-Reny\'i (ER) networks with mean degree $\langle k \rangle =8$.  

\begin{figure*}[t!]
\centering
\includegraphics[width=1.63\columnwidth]{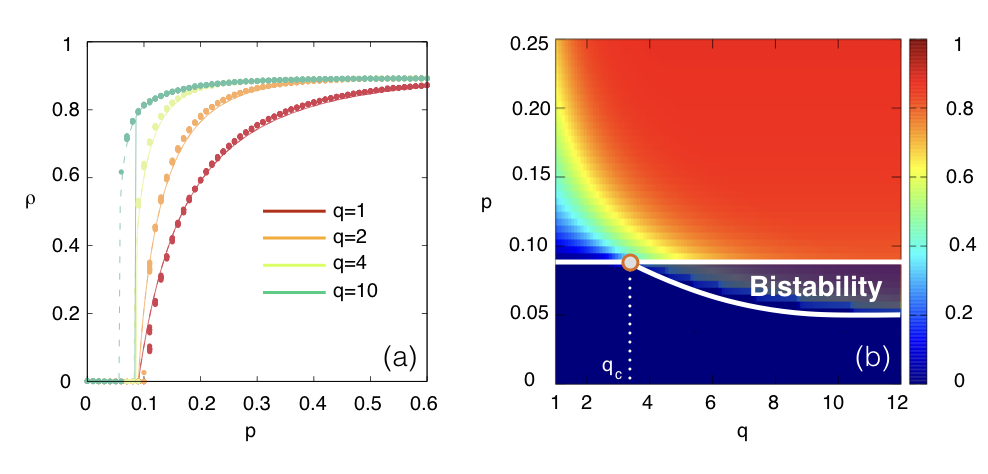}
\caption{Panel (a) shows the epidemic size $\rho$ as a function of the infectivity $p$ for several values of the interaction parameter $q$. The curves (solid and dashed) correspond to the results obtained by iterating Eqs.(\ref{Eq.II}-\ref{Eq.SI}) when performing forward (solid) and backward (dashed) continuation schemes (see text for details). In its turn, points represent the results from 50 realizations of Monte Carlo simulations for each value of $p$. In panel (b) we plot the phase diagram $\rho(q,p)$ for the cooperative case. The solid curves denote the points where epidemic onsets take place. The critical point at $q_c$ pinpoints where the bistability (faded) region appears. In both panels we use an ER network of $N=1000$ nodes with $\langle k\rangle = 8$, while $r=0.75$.}
\label{fig2}
\end{figure*}

\section{Cooperative diseases}
The spread of cooperative diseases can be captured in our formalism by setting $q\geq 1$ in Eqs.(\ref{Eq.II})-(\ref{Eq.SI}).  In this case, we first explore the effects of increasing the cooperation strength. To do so, we study the dependence of the epidemic size on the infectivity $p$ and the interaction parameter $q$. For single diseases, the epidemic size is defined as the fraction of agents infected when the epidemic has reached its stationary state. Following this definition, the epidemic size in the case of two diseases, denoted in the following as $\rho$, reads as:
\begin{equation}
\rho = \frac{1}{N}\sum_{i=1}^N\left(\rho^{IS}_i +\rho^{SI}_i +\rho^{II}_i\right)
\end{equation} 
\smallskip

In Fig.~\ref{fig2}.a we represent the epidemic diagrams (curves) for several values of $q$ which range from the non-interacting case ($q=1$) to that of large cooperation ($q=10$). Similarly to the case of single non-interaction diseases, there is a threshold $p_c$ that separates the disease-free regime and the epidemic phase. Interestingly, the shape of the transition between both solutions strongly depends on the coupling $q$ between the two SIS dynamics. Specifically, for low values of the interaction between diseases, we observe the characteristic second-order transition of the single SIS model, i.e., at the epidemic threshold the absorbing state is no longer stable and the epidemic size $\rho$ grows smoothly as the infectivity increases. However, as the cooperation is strengthened, the continuous transition turns into a discontinuous one identified by a sharp variation of the epidemic size. This abrupt transition yields a bistable region for some $p$ values in which both the epidemic and the disease-free states are simultaneously stable. This bistability is manifested by making a forward and backward continuation of in $p$ when solving Eqs.~(\ref{Eq.II})-(\ref{Eq.SI}) for a fixed value of $q$. The continuation method for the forward (backward) continuation solves Eqs.~(\ref{Eq.II})-(\ref{Eq.SI}) by using as the initial condition for a value $p \pm\delta$  the solution of these equations for $p$ perturbed with some small noise. The solutions corresponding to the forward and backward continuations as plotted as solid and dashed curves respectively.

\smallskip

To confirm the change in the nature of the epidemic onset observed from the solution of Eqs.(\ref{Eq.II})-(\ref{Eq.SI}) we perform simulations of the mechanistic model in which each agent possesses a particular dynamical state (SS, IS, SI or SS) which is updated according to the contagion network and the microscopic rules defined in Sec. II. Fig. 2.a shows the results (points) obtained from $50$ Monte Carlo simulations for each pair of ($p,q$) values. From these results, it becomes clear that our model is able to reproduce very accurately the phenomena arisen from the cooperation between both diseases. In particular, for $q=10$, the Monte Carlo solutions clearly reveal that the the transition becomes abrupt yielding the predicted bistable region in which both the epidemic and the disease-free states are simultaneously stable. 

\begin{figure}[t!]
\centering
\includegraphics[width=0.9\columnwidth]{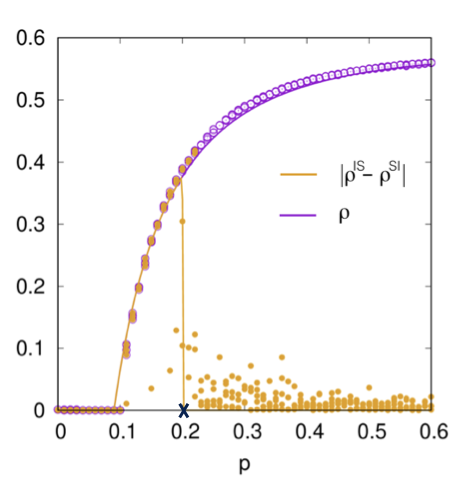}
\caption{Evolution the global epidemic size $\rho$ (purple empty points) and the difference between the prevalences of both diseases (orange solid points) as a function of the infectivity $p$ in the fully competitive case $q=0$. These points are the results obtained from 50 realizations of Monte Carlo simulations for each value of $p$. In its  turn, solid lines denote the solution obtained by iterating Eqs.~(1)-(3). Finally, the cross shows the theoretical estimation of the threshold $p_c^{\prime}$ as obtained from Eq. (\ref{pc2}). The network and the recovery probability are the same as those used in Fig.~\ref{fig2}. }
\end{figure}

\smallskip

Finally, we take advantage of the former validation and the efficiency of the Markovian formalism to fully characterize the phase diagram in the $(p,q)$ space. In Fig.~2.b we show the epidemic prevalence $\rho(q,p)$ and the curve $p_c(q)$ where epidemic onsets take place for each value of $q$. Interestingly, we can accurately identify the critical value $q_c$ for which this curve divides into two branches, signalling when bistability region (faded in the panel) shows up. This critical point ($q_c,p_c(q_c)$) pinpoints the border between three phases: the epidemic, the disease-free, and the bistable ones.

\section{Competitive diseases}

Once shown the formalism for the cooperative case, we now tackle the competition between two pathogens that propagate simultaneously. As a result of this competition, the infection probability of a pathogen decreases when affecting individuals who are already infected by the other one. Competitive interactions can be accomodated in Eqs.~(1)-(3) by setting $q<1$. This choice includes a particular case that has drawn a lot of attention \cite{int5,int6,int7,int8}: that of mutually exclusive pathogens. This particular scenario corresponds to the case $q=0$ in Eqs.~(1)-(3) and captures the situation in which the infection by one pathogen generates cross-immunity to the other one, thus making state II inaccesible to agents.
\smallskip

The theoretical study of mutually exclusive pathogens has revealed the existence of different epidemic regimes depending on the value of the contagion rate $p$ and also on the initial partition of infected seeds for both diseases. According to these studies \cite{int5,int6,int7,int8}, the outcome of fully competitive disease ranges from the dominance of one single disease (and the resulting extinction of the other one) to a regime of coexistence in which the whole epidemic prevalence contains infected individuals of both diseases.
\smallskip

To study this phenomenology in the Markovian framework we set, as anticipated above, $q=0$ and study the impact of both diseases as a function of the contagion rate $p$. Regarding the initial infectious seeds, we bias the initial configuration towards one of the pathogens by infecting $2\%$ of the population with the first disease and $1\%$ with the second one. Starting from this setup we iterate Eqs.~(1)-(3) and compute the global epidemic size $\rho$ together with the difference between the prevalences of each disease, $|\rho^{IS}-\rho^{SI}|$, where=
\begin{eqnarray}
\rho^{IS}&=& \frac{1}{N}\sum_{i=1}^N \rho^{IS}_i\;, \\
\rho^{SI}&=& \frac{1}{N}\sum_{i=1}^N \rho^{SI}_i\;. 
\end{eqnarray} 
With these two order parameters we are able to distinguish between the case of full dominance of one disease ($\rho=|\rho^{IS}-\rho^{SI}|>0$ and that corresponding to equal prevalence ($\rho>0$ and $|\rho^{IS}-\rho^{SI}|=0$).
\smallskip

In Fig.~3 we monitor these two order parameters as a function of the infection probability $p$. From this plot, it becomes clear that the Markovian framework reproduces the phenomenology previously reported for competitive diseases. Namely, for $p<p_c$, the absorbing disease-free state is the only stable solution. Note that, being the infection probabilities of both pathogens the same, this epidemic threshold is exactly the same as for independent SIS diseases. Therefore, $p_c$ is given by \cite{gomezEPL,gomezPRE}:
\begin{equation}
p_c = \frac{r}{\Lambda_{max}({\bf A})}\;,
\label{pc1}
\end{equation}
where $\Lambda_{max}({\bf A})$ is the maximum eigenvalue of the adjacency matrix ${\bf A}$.  Above this threshold, $p>p_c$ the global epidemic size grows smoothly as the infectivity $p$ increases. However, two different behaviors show up. For $p>p_c$ we find a dominance regime in which one disease (here the first one) prevails over the other one (whose prevalence is zero). In particular, the dominant disease is that with the larger initial proportion of infected individuals (here the first one). However, the dominance regime suddenly breaks up when the infectivity reaches a second threshold, denoted in the following as $p_c^{\prime}$. For $p>p_c^{\prime}$, the second pathogen no longer disappears and the steady state now comprises an equal prevalence of both diseases. Thus, the coexistence regime appears when $p>p_c^{\prime}$. These results (obtained by solving Eqs.~(1)-(3) with $q=0$) are totally in agreement with those obtained from Monte Carlo simulations of the mutually exclusive coupled SIS model, reported as points in Fig.~3.a.
\smallskip

It is possible to explain the full dominance regime in physical terms by recalling that once $p\gtrsim p_c$ the most abundant pathogen blocks, due to the cross-immunization effect, many contagion pathways for the minority one. This way, the effective network that remains for the spread of the minority pathogen has an effective epidemic threshold larger than the original one, $p_c$, thus preventing its dissemination for $p\gtrsim p_c$. However, as $p$ increases the initial unbalanced configuration loses its relevance and the transmission of both diseases become identical, thus leading to a coexistence regime. This latter phase of the dynamics shows up when the second threshold is exceeded, $p> p_c^{\prime}$.
\smallskip

Let us now take advantage of the validity of the Markovian formalism to derive some analytical calculations about the second threshold, $p_c^{\prime}$, separating the full dominance and the coexistence regimes in the case of mutually exclusive diseases ($q=0$). We start by making the following change of variables:
\begin{eqnarray}
\rho^{t}_i&=&[\rho^{IS}]^{t}_{i} +[\rho^{SI}]^{t}_{i} \;, \\
\Delta^{t}_i&=&[\rho^{IS}]^{t}_{i} -[\rho^{SI}]^{t}_{i} \;.
\end{eqnarray}
Note that this change leaves the third set of variables $[\rho^{II}]^{t}_{i}$ unaltered since for the case $q=0$ we have $[\rho^{II}]^{t}_{i}=0$ $\forall i$. From Eqs.~(1)-(2) we can write the Markovian equations governing the time evolution of the new variables as:
\begin{eqnarray}
\rho^{t+1}_i&=&(1-r)\rho^{t}_i+(1-\rho^{t}_i)q_i(\vec{\rho^t})\ ,\label{eq:rho1}\\
\Delta^{t+1}_i&=&(1-r)\Delta^{t}_i+(1-\rho^{t}_i)q_i(\vec{\rho^t})\left(f^{IS}-f^{SI}\right)\label{eq:delta2}\ ,
\end{eqnarray}
where $f^{IS}=f^{IS}(\vec{\rho^{t}},\vec{\Delta^{t}})$ and $f^{SI}=f^{SI}(\vec{\rho^{t}},\vec{\Delta^{t}})$, and we have defined:
\begin{equation} 
q_i^t(\vec{\rho^t})=\left[1-\prod_{j=1}^{N}(1-pA_{ij}\rho^{t}_j)\right]\;,
\end{equation} 
for the sake of clarity. Note that the first equation, Eq.~(\ref{eq:rho1}), is formally identical to that of an ordinary SIS model. This implies that, either in the case that one pathogen dominates over the other or in the regime in which they coexist, the overall prevalence is the same as that of one single pathogen spreading in the network. Thus, the form of 
Eq.~(\ref{eq:rho1}) pinpoints that, in order to capture the second transition point $p_c^{\prime}$, we should focus on the behavior of $\{\Delta^{t}_i\}$, Eq.~(\ref{eq:delta2}), since in the full dominance phase ($p_c<p<p_c^{\prime}$) we have $\{\Delta^{t}_i\}=\vec{0}$, whereas the coexistence phase ($p>p_c^{\prime}$) is characterized by $\{\Delta^{t}_i\}\neq\vec{0}$.
\smallskip
 
Let us consider that the dynamics is in its stationary regime, i.e. $\rho^{t+1}_i=\rho^{t}_i=\rho^{*}_i$ and $\Delta^{t+1}_i=\Delta^{t}_i=\Delta^{*}_i$ $\forall i$. In this case Eq.~(\ref{eq:delta2}) becomes:
\begin{equation} 
\Delta^{*}_i=\rho_i^*\left(f^{IS}(\vec{\rho^{*}},\vec{\Delta^{*}})-f^{SI}(\vec{\rho^{*}},\vec{\Delta^{*}})\right)\;.
\label{Deltastar}
\end{equation} 
Since we are interested in the capturing the transition between $\vec{\Delta^*}=0$ and $\vec{\Delta^*}\neq 0$ we consider that the values $\Delta_i^*$ are small, $\Delta_i^*=\epsilon_i^*$. This allow us to linearize Eq.~(\ref{Deltastar}), and obtain the equation that has to be fulfilled at $p=p_c^{\prime}$:
\begin{eqnarray} 
\epsilon_i^*&=&\sum_{l=1}^{N}\left[\rho^*_i\frac{p^{\prime}_c A_{il}q_i(\vec{\rho^*}/2)}{(1-p^{\prime}_cA_{il}\rho^*_l/2)(1-q_i(\vec{\rho^*}/2)^2)}\right]\epsilon_l^*\nonumber\\
&=&\sum_{l=1}^{N}{\cal M}_{il}(\vec{\rho^*}; p^{\prime}_c)\epsilon_l^*\;.
\label{pc2}
\end{eqnarray}
Thus, in order to find the value $p_c^{\prime}$ one needs to find the minimum value of $p$ that fulfills that matrix ${\cal{M}}$ has $1$ as eigenvalue. In practical terms, since ${\cal{M}}$ depends on the overall prevalence, $\vec{\rho^*}(p)$, one should first solve the SIS diagram $\vec{\rho^{*}}(p)$ for a single disease and, by inserting the resulting stationary values $\vec{\rho^*}(p)$ in matrix ${\cal{M}}$, identify the value $p_c^{\prime}$ that fulfils Eq.~(\ref{pc2}).
\smallskip

\begin{figure}[t!]
\centering
\includegraphics[width=0.90\columnwidth]{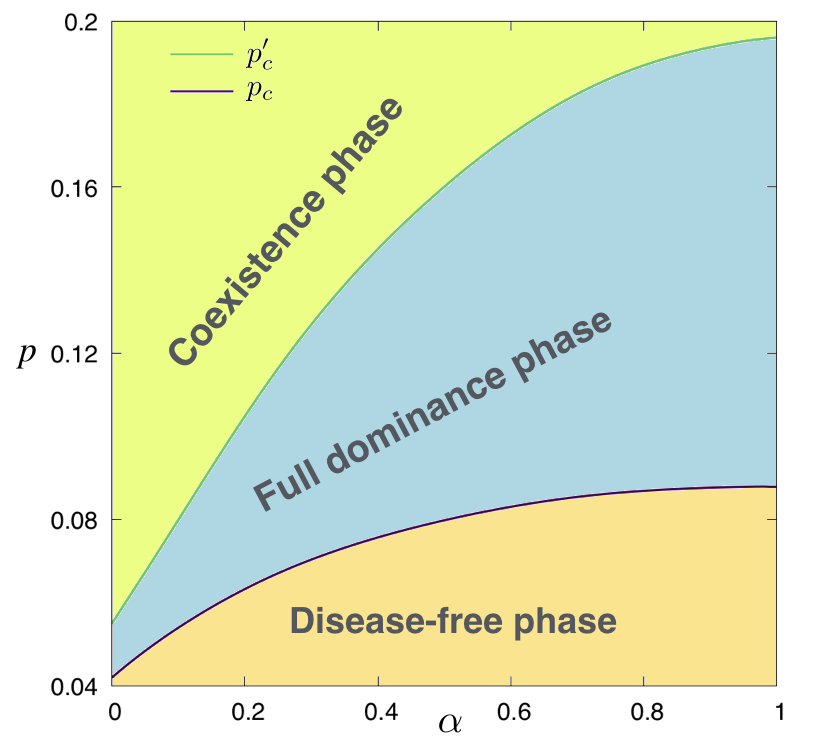}
\caption{Phase diagram of the fully-competitive model ($q=0$) in the $(\alpha,p)$ space. Parameter $\alpha$ governs the degree heterogeneity of the underlying contact network (see main text). This diagram is obtained by calculating the curves $p_c(\alpha)$ and $p_c^{\prime}(\alpha)$ obtained from Eqs.~(\ref{pc1}) and (\ref{pc2}) respectively.  These curves pinpoint the borders between the three possible regimes as reported in the plot and show that the second threshold $p_c^{\prime}$ increases as homogeneity ($\alpha$) increases, as already known for the first (epidemic) threshold $p_c$. The recovery rate has been fixed to $r=0.75$.}
\label{fig4}
\end{figure}

Following Eq.~(\ref{pc2}) we obtained the estimation of $p_c^{\prime}$ for the ER graphs used in our numerical simulations so far. The result (see the cross in Fig.~3) reveals the accuracy of the theoretical prediction. We can therefore use this result to analyze, without the need of solving Eqs.~(1)-(2), the two thresholds, $p_c$ and $p_c^{\prime}$, for any given network. In order to understand the role of network topology on the coexistence of the two diseases,  we have explored the evolution of these two thresholds in a model \cite{BAER} that allows a smooth interpolation between ER and Barb\'asi-Albert scale-free (SF) networks by changing one parameter, $\alpha\in[0,1]$, so that $\alpha=0$ corresponds to the SF limit and $\alpha=1$ corresponds to ER graphs. In Fig.~\ref{fig4} we show the curves $p_c(\alpha)$ and $p_c^{\prime}(\alpha)$ as obtained from Eqs.~(\ref{pc1}) and (\ref{pc2}) respectively. As a result we obtain the phase diagram in the $(\alpha,p)$ space showing the limits between the three phases. Interestingly, the second threshold $p_c^{\prime}$ follows a decreasing trend as heterogeneity increases ($\alpha\rightarrow 0^{+}$) similar to the well-known behavior of the epidemic threshold $p_c$.
\smallskip

\begin{figure*}[t!]
\centering
\includegraphics[width=1.6\columnwidth]{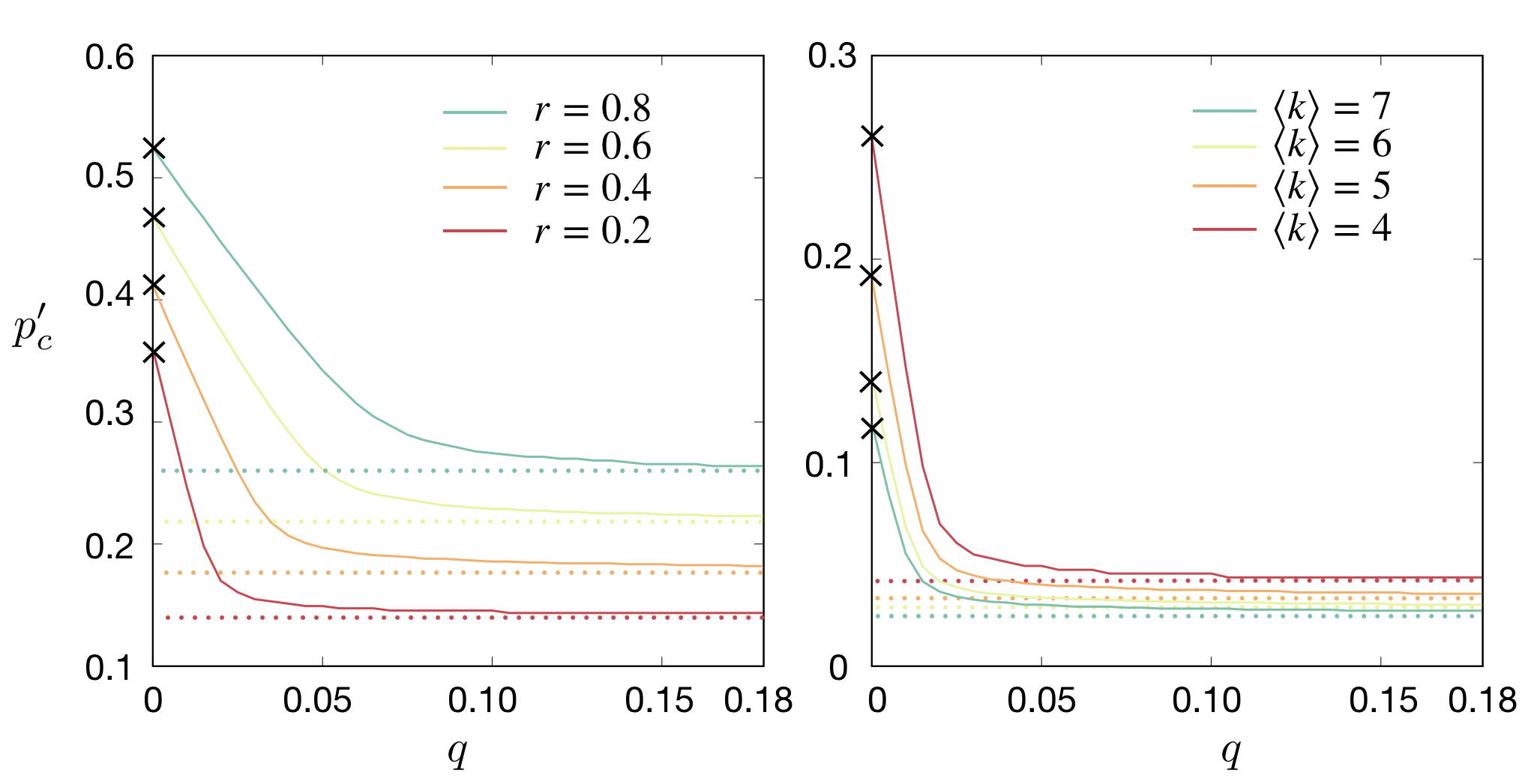}
\caption{In this figure we show the second epidemic threshold $p_c^{\prime}$  as a function of the interaction parameter $q$ for several values of the recovery rate $r$ (a) and different average degree of the underlying ER network (b). In (a) the average degree has been set to $\langle k\rangle =4$ whereas in panel (b) the recovery rate is fixed to $r=0.2$. In both panels the crosses account for the theoretical estimation of $p_c^{\prime}$ in the fully competitive case ($q=0$) according to Eq.~(\ref{pc2}). Dotted lines indicate the value of the threshold $p_c$ obtained from Eq. (\ref{pc1}) for each ($r,\langle k\rangle$).}
\end{figure*}

Finally, we use the Markovian framework, Eqs.~(1)-(3), to explore less restrictive competitive scenarios in which $q\neq 0$.  This way, in Fig. 5 we study the dependency of the second epidemic threshold $p_c^{\prime}$ on the interaction parameter $q$ for several values of the recovery rate $r$ and the average degree of the ER contact network $\langle k\rangle$. 
In both panels of Fig. 5, it becomes clear that increasing $q$ from the fully-competitive case makes the full-dominance regime more vulnerable. Note, however, that, even when the two pathogens are not mutually exclusive from a microscopical point of view ($q\neq0$), the existence of a negative interaction between them can lead to the vanishing of the weak disease in the macroscopic state. Obviously, as $q$ increases, this competition is softened and the second threshold $p_c'$ approaches the first one $p_c$ (dotted lines in Fig. 5). When $p_c'=p_c$ the full dominance phase no longer shows up. Interestingly, this effect strongly depends on the value of the recovery rate and the average degree. In particular, Fig.~5.a reveals that the larger is the recovery rate the more resilient is the full dominance regime. This can be explained by noticing that increasing the recovery rate prevents the formation of large clusters of the dominant disease, thus making easier the propagation of the other one. In its turn, in Fig. 5.b we report that increasing the average degree of the network (while keeping constant the recovery rate) favors the emergence of the coexistence regime. To explain this, let us remark that the extinction of one disease occurs since the other one blocks many of its spreading pathways. Thus, a larger average degree promotes more potential contagion routes, making the full dominance regime more vulnerable.
\smallskip

\section{Conclusions}

Phenomena such as the existence of simultaneous outbreaks or the extinction of some infectious strains due to the presence of other ones demand the incorporation of the interaction between different pathogens in epidemic models. The most relevant contribution of this work is to provide a versatile Markovian framework capable of capturing different types of interaction between co-existing diseases. In particular, the scenarios that can be addressed by this formalism range from a strong cooperation between pathogens to the mutually exclusive case in which the infection by one pathogen automatically creates cross-immunity to the others. Besides, this framework accounts for the whole structure of connections of the underlying contact network, thus abandoning the assumptions about the statistical equivalence of nodes within the same degree class. The validity of the proposed equations has been tested by comparing with the results obtained from Monte Carlo simulations, showing an excellent agreement for any degree of epidemic prevalence.
\smallskip

Supported by the validity of the Markovian equations, we have explored the role that the interaction between diseases plays on their unfolding. For cooperative diseases, we have shown that, as the interaction between pathogens increases, the smooth transition at the epidemic thresholds turns into a first-order one in which there is an abrupt transition between the absorbing (healthy) state and the epidemic one. 
We have used the formalism to compute the critical value for the degree of cooperation, $q_c$, that  triggers the abrupt transition. In its turn, for competitive diseases we have analyzed the case of mutually exclusive pathogens.  In this case we have shown that the epidemic phase is divided into two different regimes: the full dominance phase (in which only one pathogen spreads across the network while the other disappears) and the coexistence phase (in which both pathogens spread simultaneously). Based on the linearization of the Markovian equations, we have derived an analytical estimation of the infectivity threshold, $p_c^{\prime}$, that separates both phases. Finally, we have studied how the full dominance phase disappears as competition decreases by monitoring how the former threshold approaches the epidemic one, $p_c$. 
\smallskip

In a nutshell, the Markovian benchmark presented here has allowed a systematic study of different degrees of positive and negative interactions between the pathogens and the accurate characterization of the corresponding phase diagrams. The analytical derivation of the infection threshold between full-dominance and coxistence regime for any arbitrary network opens the door to the design of contention measures via the introduction of highly infective but innocuous computer viruses aimed at decreasing the damage of malware \cite{virus}. Finally, our formalism paves the way for the study of more sophisticated interacting spreading processes such as complex social contagions \cite{complex0,complex1,complex2,simplicial} applied, for instance, to the competition of ideas \cite{NJP} or innovations \cite{int9} .

\acknowledgments  We acknowledge inspiring discussions and useful insights from L.M. Flor\'{\i}a and Y. Moreno.  D.S.-P. acknowledges financial support from Gobierno de Arag\'on through a doctoral fellowship. J.G.-G and D.S.-P. acknowledge support from MINECO (Grants FIS2015-71582-C2 and FIS2017-87519-P), and Gobierno de Arag\'on/Fondo Europeo de Desarrollo Regional (FEDER) (Grant E36-17R to FENOL group). FGh acknowledges the particial support by the Deutsche Forschungsgemein-schaft under grant GH 176/1-1, within the idonate program (project 345463468). SM acknowledges partial financial support from the Agencia Estatal de Investigacion (AEI, Spain) and Fondo Europeo de Desarrollo Regional under Project PACSS Project No. RTI2018-093732-B-C22  (MCIU, AEI/FEDER,UE) and through the Mar\'ia de Maeztu Program for units of Excellence in R\&D (MDM-2017-0711).

\end{document}